\documentclass[10pt,conference]{IEEEtran}

\usepackage{xspace}
\usepackage{longtable}
\usepackage{array}
\usepackage[table]{xcolor}
\usepackage{url}

\usepackage{color}
\usepackage{nameref}
\usepackage{multirow}
\usepackage{tabularx}
\usepackage{censor}
\usepackage{enumitem}
\usepackage[most]{tcolorbox}
\usepackage{mdframed}
\usepackage{fontawesome5}

\definecolor{lightblue}{RGB}{0,0,100}

\newtcolorbox{MyBox}{
  colback=white,
  colframe=lightblue,
  fonttitle=\bfseries,
  coltitle=black,
  sharp corners,
  boxrule=1pt,
  left=5pt,
  right=5pt,
  top=5pt,
  bottom=5pt,
  breakable
}

\newmdenv[
  backgroundcolor=black!6,
  leftline=true,
  rightline=false,
  topline=false,
  bottomline=false,
  linecolor=black,
  linewidth=3pt,
  innerleftmargin=10pt,
  innerrightmargin=10pt,
  innertopmargin=8pt,
  innerbottommargin=8pt,
  skipabove=8pt,
  skipbelow=8pt
]{rqbox}

\newtcolorbox{databox}{
  colback=teal!20!white,
  boxrule=0pt,
  enhanced,
  borderline west={4pt}{0pt}{black},
  left=10pt,
  right=10pt,
  top=8pt,
  bottom=8pt,
  sharp corners
}

\newtcolorbox{resultbox}{
  enhanced,
  colback=blue!5,
  boxrule=0pt,
  sharp corners,
  borderline west={4pt}{0pt}{black},
  left=10pt,
  right=10pt,
  top=8pt,
  bottom=8pt
}

\begin{document}

\title{Exploring Quantum Software Testing Across Research and Practice: Emerging Results from a Multivocal Literature Review}

\author{
\IEEEauthorblockN{Rodolfo Gil-Pereira}
\IEEEauthorblockA{\textit{University of Calgary} \\
Calgary, Canada \\
rodolfo.gilpereira@ucalgary.ca} \\

\IEEEauthorblockN{Italo Santos}
\IEEEauthorblockA{\textit{University of Hawaii at Manoa} \\
United States \\
isantos3@hawaii.edu}

\and

\IEEEauthorblockN{Ronnie de Souza Santos}
\IEEEauthorblockA{\textit{University of Calgary} \\
Calgary, Canada \\
ronnie.desouzasantos@ucalgary.ca} \\

\IEEEauthorblockN{Cleyton Magalhães}
\IEEEauthorblockA{\textit{UFRPE} \\
Brazil \\
cleyton.vanut@ufrpe.br}
}
\maketitle

\begin{abstract}
This paper presents preliminary findings from a multivocal literature review investigating how quantum software testing is characterized across academic and practitioner-oriented sources. Our study integrated peer-reviewed studies with gray literature, including blogs, tutorials, forums, technical reports, documentation pages, and company webpages. Our results indicate a rapidly evolving but fragmented ecosystem involving classical adapted testing approaches, quantum-specific techniques, statistical validation methods, simulators, debugging environments, and verification frameworks. The reviewed material also revealed recurring challenges related to scalability limitations, hardware noise, probabilistic execution, limited observability, and immature tooling ecosystems. These findings provide an initial characterization of how research and practice currently discuss quantum software testing challenges, techniques, and tooling.
\end{abstract}


\section{Introduction}
\label{sec:introduction}

Quantum Software Engineering (QSE) investigates the application of software engineering principles, methods, and tools to the development of quantum software systems~\cite{zhao2020quantum,piattini2021toward}. As quantum programming frameworks and cloud-accessible quantum hardware become increasingly available, there is growing interest in engineering practices that support the development of reliable and maintainable quantum software~\cite{piattini2021toward}. Within this context, Quantum Software Testing (QST) has emerged as a key activity for assessing the correctness, reliability, and quality of quantum programs and hybrid quantum-classical systems~\cite{ali2022software,oldfield2025faster}.

Testing quantum software presents challenges that differ from those of classical software because of the probabilistic nature of quantum computation, hardware noise, limited observability, and scalability constraints~\cite{ali2022software,oldfield2025faster,murillo2025quantum}. As a result, a broad range of testing techniques, tools, and practices has emerged to address these challenges~\cite{garcia2023quantum,oldfield2025faster,murillo2025quantum}. However, current knowledge remains dispersed across the growing body of research and practitioner discussions on QST, making it difficult to obtain a comprehensive view of current testing practices, challenges, and tooling~\cite{garcia2023quantum,murillo2025quantum,zappin2025challenges}. To address this gap, we conducted a multivocal literature review integrating evidence from academic and gray literature. Our study investigates the following research question:

\begin{quote}
\textbf{RQ. How do research literature and practitioner discussions characterize QST practices, challenges, and tooling?}
\end{quote}

We synthesized evidence from 110 academic and practitioner sources to characterize testing techniques, challenges, and tooling discussed across the current QST ecosystem. The results provide a consolidated view of the state of QST and identify recurring technical and ecosystem challenges that can inform future research and practice.
\section{Background and Related Work}
\label{sec:background}

Quantum Software Testing includes activities intended to support validation, fault detection, debugging, and quality assurance in quantum software systems~\cite{zhao2020quantum, ali2022software}. Existing literature characterizes QST as spanning multiple stages of the quantum software lifecycle, including test generation, execution, adequacy assessment, verification, and debugging~\cite{zhao2020quantum}. In practice, testing workflows are commonly centered around the execution of quantum circuits through simulators or cloud accessible quantum hardware using development ecosystems such as Qiskit, Cirq, Q\#, ProjectQ, OpenQASM, and pyQuil~\cite{zhao2020quantum, ali2022software}. These workflows frequently combine classical control logic with quantum execution pipelines, reflecting the hybrid nature of quantum software systems~\cite{ali2022software, piattini2021toward}. Existing studies also indicate that repeated executions and statistical analysis are commonly required to evaluate output distributions and program behavior under probabilistic execution conditions~\cite{oldfield2025faster}. 

Recent research has proposed multiple testing approaches specifically tailored to quantum programs. Existing work discusses strategies such as metamorphic testing, mutation testing, fuzz testing, differential testing, specification based testing, and assertion based verification~\cite{zhao2020quantum, oldfield2025faster}. Several tools and frameworks have also been introduced to operationalize these techniques. MorphQ supports metamorphic testing through the generation and transformation of quantum programs to identify inconsistencies in quantum computing platforms~\cite{paltenghi2023morphq}. QuanFuzz proposes a fuzz testing strategy focused on generating quantum sensitive inputs and improving test coverage for quantum programs~\cite{wang2018quanfuzz}. Muskit introduces mutation operators and tooling support for evaluating the effectiveness of test suites in Qiskit based programs~\cite{mendiluze2021muskit}. Existing literature also reports increasing interest in debugging support, assertion libraries, probabilistic verification, and adequacy criteria adapted to quantum execution environments~\cite{zhao2020quantum}. 

Despite these advances, prior work suggests that QST tooling and engineering practices remain relatively fragmented~\cite{piattini2021toward, zappin2025challenges}. Existing studies report recurring difficulties associated with unstable APIs, rapidly evolving frameworks, hardware variability, limited debugging support, and the absence of mature benchmark repositories and standardized testing processes~\cite{mendiluze2021muskit, zappin2025challenges}. Furthermore, much of the current evidence regarding QST practices remains concentrated in academic proposals, while less is known about how practitioners discuss, adapt, and operationalize testing activities in real-world quantum software development contexts.
\vspace{-5pt}
\section{Method}
\label{sec:method}
We employed a multivocal literature review to investigate how QST is characterized across academic and practitioner-oriented sources. Multivocal literature reviews combine evidence from peer-reviewed studies and gray literature, enabling the synthesis of complementary perspectives from research and practice~\cite{garousi2019guidelines,neto2019multivocal}. This approach is particularly suitable for emerging areas such as QST, where tooling ecosystems, tutorials, documentation, blogs, and practitioner discussions frequently evolve faster than formal academic publication cycles~\cite{garousi2016need,garousi2019guidelines}. Consistent with the objective of a multivocal review, our goal was to synthesize evidence from both academic and practitioner-oriented sources to develop a comprehensive characterization of QST rather than to compare these evidence sources separately. We integrated academic studies on QST techniques, tools, and challenges with practitioner-oriented sources (e.g., blogs, tutorials, documentation, and forums), following established guidelines for search, selection, extraction, and synthesis~\cite{garousi2019guidelines,neto2019multivocal}. \\

\noindent \textbf{Search Strategy.} The search for both academic and gray literature was conducted using Google Search (google.com) between February and April 2026. The search process was conducted in two phases. First, we conducted a broad search to capture a wide range of potentially relevant sources on software, quantum computing, and testing.\footnote{\textit{SOFTWARE AND QUANTUM AND TESTING}} This phase considered the top 100 Google search results.

In the second phase, we refined the search strategy using ten focused queries targeting specific aspects of QST.\footnote{\textit{Quantum software testing}; \textit{Quantum software AND testing practices}; \textit{Quantum program testing examples}; \textit{Challenges quantum software testing}; \textit{Quantum testing AND debugging}; \textit{Quantum software testing AND noise}; \textit{Quantum software testing AND simulation}; \textit{Quantum software testing AND verification}; \textit{Quantum software testing tools}; \textit{Limitations quantum software testing}} These queries were designed to retrieve gray literature discussing testing workflows, debugging activities, simulation, tooling, verification, limitations, and testing challenges. For each focused query, the top 25 Google search results were extracted, resulting in 250 additional URLs. Combined with the initial search, the overall process produced a dataset of 350 URLs. The academic literature component focused on studies discussing QST methodologies, tools, frameworks, and challenges, including surveys, roadmap papers, methodological proposals, and empirical investigations. The gray literature component focused on practitioner-oriented materials that discuss how QST is operationalized in practice, including webpages, technical blogs, tutorials, documentation pages, company resources, white papers, forums, and practitioner discussions. \\

\noindent \textbf{Selection Process and Data Extraction.} We applied inclusion (IC) and exclusion criteria (EC) after each URL was visited and manually reviewed. URLs were not removed from the dataset during the search process. Instead, sources that did not satisfy the IC were retained in the spreadsheet but marked as ``not identified'' during extraction when relevant information could not be obtained. The IC was IC1: academic and gray literature discussing QST. The IC was: IC1 - academic and gray literature discussing QST. The EC were as follows: EC1 - unavailable or inaccessible sources; EC2 - incomplete texts or partial materials; EC3 - non-English sources; EC4 - duplicated URLs; and EC5 - sources without substantive technical or practitioner-oriented content related to QST. 

The extraction process was conducted in two stages. First, we extracted descriptive metadata from each source, including title, source type, platform, and publication year. Second, we extracted information directly related to the research question, including testing techniques, testing challenges, testing tools or frameworks, testing recommendations, testing targets, and quality attributes addressed during testing activities. To support this process, GPT-5.5 Instant was used to analyze each source individually and identify relevant excerpts associated with the extraction categories defined in the study, following emerging recommendations for the use of LLMs in evidence synthesis and literature reviews~\cite{felizardo2024chatgpt}. The model was instructed to return only exact quotations identified in the analyzed material.\footnote{\textit{Read the uploaded material and identify excerpts discussing testing challenges in QST. Return only exact quotations from the source. If no relevant information is identified, return ``not identified''.}} A separate ChatGPT session was used for each URL to reduce contextual contamination across sources. For webpages, the content was included directly in the prompt. To reduce the risk of hallucinated or inaccurate extractions, every excerpt generated by the LLM was manually verified against the original source material by two authors before being included in the dataset. Disagreements were resolved through discussion until consensus was reached. \\

\noindent \textbf{Data Analysis and Synthesis.} The extracted data were synthesized using descriptive and thematic analysis procedures~\cite{cruzes2011recommended}. Descriptive analysis was used to summarize publication characteristics, source types, recurring tools, and commonly discussed testing activities across the collected sources. Thematic synthesis was then used to identify recurring patterns related to testing practices, challenges, tooling ecosystems, and practitioner recommendations. Codes extracted from the sources were iteratively grouped into broader analytical themes, enabling comparison between academic discussions and practitioner-oriented gray literature.

\section{Findings} 
\label{sec:findings}

Our review resulted in 110 sources discussing QST across academic and practitioner-oriented materials. The dataset included 57 peer-reviewed studies and 53 gray literature sources distributed across multiple formats, including 19 practitioner blogs, 14 company webpages, 6 documentation pages and related technical documents, 5 discussion forums, 3 tutorials, and 6 technical reports. The temporal distribution of the dataset suggests that interest in QST has intensified substantially in recent years. Although the publication year could not be reliably identified for 21 sources, most identified publications were concentrated between 2024 and 2025, with 28 sources published in 2025 and 22 in 2024. Earlier years presented progressively smaller numbers of publications, including 8 sources from 2023, 7 from 2022, 9 from 2021, 5 from 2020, 3 from 2019, 2 from 2018, and 1 from 2015. Overall, the distribution indicates the recent growth and rapidly evolving nature of QST discussions across both research and practitioner communities.
\vspace{-5pt}
\subsection{Testing Techniques}

We identified five major categories of testing techniques discussed across the reviewed material. These categories reflect different strategies adopted to address the probabilistic behavior, state complexity, noise sensitivity, and execution constraints associated with quantum software systems as described below:

\begin{table*}[!htbp]
\caption{Testing Techniques}
\label{tab:testing-techniques}
\scriptsize
\renewcommand{\arraystretch}{1.2}
\begin{tabularx}{\textwidth}{p{2cm}|p{4.2cm}|p{4.2cm}|X}
\hline
\textbf{Category} & \textbf{Definition} & \textbf{Example Techniques} & \textbf{Papers} \\
\hline \hline

\textbf{Classical Adapted Techniques} &
Techniques originating from classical software testing and adapted to address the probabilistic and quantum-specific characteristics of quantum software. &
Mutation testing, fuzz testing, property-based testing, combinatorial testing, search-based testing, regression testing, unit testing, coverage criteria &
QSE001, QSE003, QSE004, QSE009, QSE010, QSE012, QSE018, QSE022, QSE024, QSE029, QSE036, QSE044, QSE045, QSE051, QSE053, QSE055, QSE060, QSE065, QSE066, QSE071, QSE077, QSE081, QSE085, QSE093 \\
\hline

\textbf{Quantum Specific Techniques} &
Techniques specifically designed for quantum computing characteristics such as superposition, entanglement, noise, and quantum state manipulation. &
Quantum state tomography, Pauli-string testing, swap test, projection-based assertions, quantum error correction, noise-aware testing, entanglement analysis &
QSE005, QSE015, QSE019, QSE021, QSE025, QSE027, QSE032, QSE037, QSE047, QSE052, QSE067, QSE069, QSE075, QSE078, QSE089, QSE091 \\
\hline

\textbf{Statistical Techniques} &
Techniques that rely on repeated executions and statistical analysis of output distributions to assess correctness and reliability. &
Chi-square test, Kolmogorov--Smirnov test, statistical assertions, hypothesis testing, statistical oracles, repeated measurements &
QSE001, QSE003, QSE011, QSE015, QSE018, QSE019, QSE022, QSE028, QSE045, QSE053, QSE061, QSE063, QSE071, QSE075, QSE077, QSE082, QSE087, QSE095 \\
\hline

\textbf{Debugging and Verification Techniques} &
Techniques focused on identifying faults, validating correctness, analyzing program behavior, or formally verifying quantum software properties. &
Debuggers, breakpoints, assertions, Hoare logic, formal verification, equivalence checking, theorem proving, static analysis &
QSE004, QSE008, QSE009, QSE011, QSE012, QSE044, QSE046, QSE050, QSE054, QSE055, QSE070, QSE072, QSE083, QSE086, QSE090, QSE094, QSE097, QSE107 \\
\hline

\textbf{Simulation Based Techniques} &
Techniques that use classical or hybrid simulators to execute, analyze, and validate quantum programs before deployment on real hardware. &
Quantum simulators, noisy simulation, hardware simulators, full-state simulation, local simulators, hardware-in-the-loop testing &
QSE006, QSE009, QSE012, QSE013, QSE026, QSE031, QSE033, QSE035, QSE043, QSE058, QSE073, QSE076, QSE079, QSE086, QSE092, QSE096, QSE097, QSE103, QSE104 \\
\hline \hline

\end{tabularx}
\end{table*}

\begin{itemize}
    \item \textbf{Classical Adapted Techniques:} correspond to testing approaches originally proposed for classical software systems and later adapted to the quantum domain. These techniques are characterized by the reuse of established software testing principles, including test generation, coverage analysis, fault detection, and behavioral validation. In the reviewed studies, these approaches were frequently used to structure testing workflows and automate the generation and execution of test cases for quantum programs.

    \item \textbf{Quantum Specific Techniques:} include approaches explicitly designed around quantum mechanical properties and quantum hardware characteristics. These techniques are characterized by the direct consideration of concepts such as superposition, entanglement, quantum measurements, reversibility, and noise. The reviewed studies used these approaches to validate quantum states, assess quantum circuit behavior, and evaluate correctness under quantum execution conditions that cannot be directly addressed through classical testing strategies.

    \item \textbf{Statistical Techniques:} comprise approaches based on probabilistic reasoning, repeated execution, and statistical analysis of quantum program outputs. These techniques are characterized by the use of statistical inference to determine whether observed execution distributions are consistent with expected program behavior. Since quantum programs produce nondeterministic outputs, the reviewed studies commonly relied on repeated measurements, sampling procedures, probability comparisons, and statistical hypothesis testing to assess correctness and reliability.

    \item \textbf{Debugging and Verification Techniques:} include approaches focused on analyzing program behavior, identifying faults, validating execution correctness, and verifying software properties. These techniques are characterized by activities related to runtime inspection, assertion checking, fault localization, logical reasoning, and formal analysis. In the reviewed studies, these approaches were used to support correctness validation and identify inconsistencies during quantum program execution.

    \item \textbf{Simulation Based Techniques:} correspond to approaches that rely on simulated environments to execute and analyze quantum software. They are characterized by the use of classical or hybrid infrastructures capable of reproducing the behavior of quantum systems without requiring direct access to physical quantum hardware. The reviewed studies frequently employed simulation based approaches to prototype quantum programs, evaluate execution behavior under controlled conditions, analyze noise effects, and support testing activities in scenarios where real hardware access was limited or unstable.
\end{itemize}

Table~\ref{tab:testing-techniques} summarizes the identified categories, their definitions, representative examples of techniques, and the corresponding studies associated with each category. Only the material that explicitly contained testing, verification, validation, debugging, statistical, or simulation-related techniques was mapped. Sources were not mapped when they did not identify a testing technique, focused primarily on modeling or infrastructure, or discussed broader quantum computing concepts without presenting a concrete testing, verification, or validation technique.
\vspace{-5pt}
\subsection{Quantum Testing Challenges}

\begin{table*}[!htbp]
\caption{Testing Challenges}
\label{tab:testing-challenges-summary}
\scriptsize
\renewcommand{\arraystretch}{1.2}
\begin{tabularx}{\textwidth}{p{2cm}|p{2cm}|p{3cm}|X}
\hline
\textbf{Category Group} & \textbf{Definition} & \textbf{Example Challenges} & \textbf{Sources} \\
\hline \hline

\textbf{Primary Challenges} &
Most commonly reported testing limitations. &
Scalability, hardware noise, destructive measurements, probabilistic outputs, debugging limitations &
QSE001, QSE003, QSE008, QSE009, QSE011, QSE012, QSE013, QSE015, QSE017, QSE018, QSE019, QSE020, QSE021, QSE022, QSE023, QSE025, QSE026, QSE028, QSE029, QSE030, QSE031, QSE034, QSE036, QSE037, QSE038, QSE041, QSE044, QSE045, QSE049, QSE050, QSE051, QSE052, QSE053, QSE055, QSE059, QSE060, QSE061, QSE062, QSE063, QSE064, QSE065, QSE066, QSE067, QSE068, QSE070, QSE071, QSE072, QSE073, QSE075, QSE076, QSE077, QSE079, QSE081, QSE082, QSE083, QSE084, QSE085, QSE086, QSE087, QSE088, QSE089, QSE091, QSE093, QSE095, QSE096, QSE097, QSE098, QSE099, QSE104, QSE105, QSE107, QSE108, QSE109 \\
\hline

\textbf{Secondary Challenges} &
Recurring constraints affecting testing activities. &
Oracle construction, hardware constraints, computational cost, immature ecosystems, skill gaps &
QSE001, QSE003, QSE006, QSE010, QSE011, QSE014, QSE015, QSE017, QSE018, QSE019, QSE020, QSE021, QSE022, QSE023, QSE026, QSE027, QSE028, QSE029, QSE032, QSE033, QSE037, QSE040, QSE041, QSE044, QSE046, QSE047, QSE048, QSE049, QSE050, QSE054, QSE055, QSE062, QSE063, QSE065, QSE066, QSE067, QSE068, QSE069, QSE070, QSE071, QSE072, QSE073, QSE074, QSE078, QSE079, QSE081, QSE082, QSE084, QSE085, QSE086, QSE087, QSE090, QSE092, QSE093, QSE094, QSE097, QSE101, QSE103, QSE106, QSE107, QSE108, QSE109 \\
\hline

\textbf{Less Frequent Challenges} &
Specialized or emerging testing concerns. &
Error correction, hardware adaptation, test input generation, formal verification &
QSE005, QSE024, QSE025, QSE029, QSE030, QSE035, QSE045, QSE046, QSE055, QSE059, QSE060, QSE075, QSE080, QSE093, QSE102, QSE107 \\
\hline \hline

\end{tabularx}
\end{table*}

We identified a broad set of challenges that make QST difficult or limit the effectiveness of testing activities, illustrated in Table~\ref{tab:testing-challenges-summary}. The most frequently reported challenges were related to scalability, hardware noise, measurement limitations, probabilistic outputs, debugging limitations, and tool limitations. The reviewed material repeatedly described how the exponential growth of quantum state spaces limits the scalability of testing and simulation techniques, while noise and decoherence reduce the reliability of execution results. In addition, the destructive nature of quantum measurements and the probabilistic behavior of quantum programs make it difficult to inspect intermediate states, reproduce executions, and determine whether failures are caused by faults or hardware instability. Multiple sources also emphasized that existing testing and debugging tools remain limited, fragmented, or poorly integrated with established software engineering infrastructures.

Secondary challenges included the oracle problem, hardware constraints, computational cost, ecosystem maturity, skill gaps, noise modeling, and noise simulation. Multiple sources discussed how defining expected outputs for quantum programs is often infeasible, particularly for probabilistic or computationally expensive algorithms. Other sources reported that restricted hardware access, limited qubit counts, short coherence times, and high execution costs constrain the feasibility of testing activities. The reviewed material also highlighted the immaturity of the current quantum software ecosystem, including fragmented tooling, a lack of standards, limited testing guidance, and insufficient expertise required to conduct QST activities effectively. Several sources additionally emphasized that developers and testers often lack the specialized knowledge necessary to understand quantum behavior, interpret probabilistic results, and adapt classical software testing practices to quantum systems.

Less frequently reported challenges included error correction, error mitigation, formal verification, quantum-specific bugs, hardware adaptation, test input generation, state observability, hardware cost, reliability, and test planning. Although discussed less often, these challenges also impose important limitations on testing activities, especially in specialized contexts such as fault-tolerant quantum computing, hardware-aware optimization, and automated test generation. Some sources also discussed the difficulty of adapting testing strategies to heterogeneous hardware platforms and identifying defects that emerge specifically from quantum behavior and execution environments.

\begin{table*}[!htbp]
\caption{Quantum software testing tools and frameworks}
\label{tab:testing-tools}
\scriptsize
\renewcommand{\arraystretch}{1.2}
\begin{tabularx}{\textwidth}{p{4cm}|X}
\hline

\textbf{Category} & \textbf{Representative Tools / Frameworks} \\
\hline \hline

\textbf{Classical Adapted Techniques} &
Muskit, QMutPy, QuMu, MutTG, Quito, QuCAT, QuSBT, QuanFuzz, NovaQ, qATG, QuraTest, QSharpCheck, QuCheck, QSharpTester, MorphQ, QDiff, QITE, Quantum Concolic Testing \\
\hline

\textbf{Quantum Specific Techniques} &
QOPS, randomized benchmarking, gate set tomography, quantum volume, mirror benchmarking framework, Virtual Distillation \\
\hline

\textbf{Statistical Techniques} &
QOIN, ZNE, Q-LEAR, QLEAR, Mitiq, CDR, L-PEC, ML-QEM, QRAFT, Probabilistic Error Cancellation \\
\hline

\textbf{Debugging and Verification Techniques} &
QDebug, Proq, QChecker, QSmell, LintQ, Giallar, Gleipnir, QSynth, Entang$\lambda$e, MQT-quSAT, CBMC, Seahorn, PySMT, Z3, Coq, Isabelle, theorem proving, model checking \\
\hline

\textbf{Simulation Based Techniques} &
Qiskit Aer, AerSimulator, QuTiP, qHiPSTER, IQS, QVM, LocalSimulator, MIMIQ, Qaptiva, QuantumSimulator, ToffoliSimulator, Cirq Simulator \\
\hline

\textbf{Supporting Testing Infrastructure} &
Bugs4Q, QBugs, MQTBench, Metriq, Quantum Benchmark, Benchpress, Quantum Testing Framework (QTF), QuTAF \\
\hline \hline

\end{tabularx}
\end{table*}

\vspace{-5pt}
\subsection{QST Tools and Frameworks}

We identified a broad ecosystem of tools, frameworks, infrastructures, and supporting environments associated with QST activities across the reviewed material. The identified tooling ecosystem included mutation testing frameworks, combinatorial and search-based testing tools, property-based testing approaches, debugging and verification environments, statistical validation and noise-mitigation techniques, simulators, benchmark repositories, and testing automation infrastructure. Across the studies, these tools were used to support activities such as automated test generation, fault detection, correctness validation, runtime inspection, probabilistic analysis, and simulation-based evaluation of quantum software systems.

The identified tools were categorized according to the primary testing activity explicitly described in the reviewed sources. The classification followed the same analytical structure as the testing technique categories presented previously, allowing the tooling ecosystem to be interpreted in terms of the dominant testing activities discussed in the literature. Only tools explicitly associated with testing, debugging, verification, statistical validation, or simulation-related activities were mapped. General-purpose quantum programming frameworks, cloud platforms, development environments, and hardware infrastructures were not categorized unless the reviewed material explicitly described their use in supporting testing-related activities. Table~\ref{tab:testing-tools} summarizes the identified categories of tools and frameworks together with representative examples extracted from the reviewed sources.

Several additional tools, frameworks, platforms, and programming environments were also identified throughout the reviewed material, including broader quantum software ecosystems and execution infrastructures such as Qiskit, Cirq, Q\#, Amazon Braket, Azure Quantum, PennyLane, Forest, ProjectQ, OpenQASM, and IBM Quantum services. However, these tools were not included in Table~\ref{tab:testing-tools} because the reviewed sources did not provide sufficiently explicit information about their direct roles in testing, debugging, verification, statistical validation, or simulation-based testing. In multiple cases, these tools were discussed primarily as execution environments, programming ecosystems, or supporting infrastructures rather than as testing specific techniques or frameworks.

\section{Discussion}
\label{sec:discuss}
This section discusses the findings in relation to previous research and presents their implications for quantum software engineering research and practice.

\subsection{Comparing Findings}

Our findings are consistent with previous research characterizing QST as a domain centered on probabilistic execution, statistical validation, debugging, verification, and the adaptation of classical software testing techniques to quantum programs~\cite{zhao2020quantum,ali2022software,oldfield2025faster}. Similarly, our review confirms earlier reports identifying hardware noise, limited observability, scalability constraints, and oracle construction as recurring technical challenges in testing quantum software~\cite{piattini2021toward,zappin2025challenges}. Our findings also align with previous studies describing a growing ecosystem of testing techniques and supporting tools, including mutation testing, fuzz testing, metamorphic testing, simulation environments, debugging support, and probabilistic verification~\cite{zhao2020quantum,oldfield2025faster,paltenghi2023morphq,wang2018quanfuzz,mendiluze2021muskit}. Likewise, we observed continued reliance on hybrid quantum classical workflows and simulator based testing, reflecting the limited accessibility and operational constraints of current quantum hardware~\cite{ali2022software,piattini2021toward}.

However, our review extends previous work by synthesizing evidence from both academic and practitioner-oriented sources. While existing studies have primarily emphasized testing techniques, technical challenges, and individual tools, our synthesis provides a broader characterization of the current QST ecosystem. In addition to established technical challenges, the reviewed material consistently identified fragmented tooling ecosystems, limited testing guidance, insufficient integration with established software engineering infrastructures, hardware accessibility constraints, and workforce skill gaps as recurring barriers to the adoption of QST practices. Our review also consolidates a broader set of testing tools, benchmark repositories, automation infrastructure, debugging environments, simulators, and noise mitigation frameworks than previously synthesized in a single study, providing a more comprehensive characterization of the current state of QST research and practice.

\subsection{Implications}
From a \textbf{\textit{research perspective}}, as expected from a multivocal literature review, our synthesis identifies several opportunities for future research. The reviewed material suggests the need for integrated testing ecosystems that combine test generation, execution, debugging, verification, benchmarking, and statistical analysis within unified development workflows. The fragmented nature of the current tooling ecosystem also indicates opportunities to develop interoperable testing infrastructures, reusable benchmark repositories, standardized testing processes, and evaluation procedures that facilitate comparison across tools and studies. Furthermore, the limited evidence on industrial testing practices suggests the need for empirical investigations examining how practitioners test quantum software in real development environments, how organizations adopt existing testing techniques, and how software engineering practices can be adapted to quantum software development. Finally, the recurring workforce and skill related challenges indicate opportunities to investigate educational strategies, professional training, and developer support for QST. From a \textbf{\textit{industry perspective}}, our findings indicate that organizations, development teams, and practitioners currently operate within a still evolving testing ecosystem. The reviewed material reports challenges related to hardware accessibility, fragmented tools, limited debugging support, insufficient testing guidance, and the absence of standardized testing processes. Organizations adopting quantum technologies should therefore establish testing strategies early in the development lifecycle, combine complementary testing techniques rather than relying on a single approach, and make systematic use of simulation environments before executing software on physical quantum hardware. Development teams may also benefit from adopting reusable testing assets, documenting testing workflows, and investing in workforce preparation as quantum software engineering practices continue to mature. More importantly, our findings suggest that improvements in tooling interoperability, testing guidance, and professional training are likely to support broader and more reliable adoption of QST.

\subsection{Threats to Validity}
This study is subject to limitations commonly associated with multivocal literature reviews. For example, the gray literature search relied on Google Search results, which may vary over time, across geographic regions, and according to search engine ranking mechanisms, while gray literature itself varies in technical depth and credibility. To mitigate these threats, we employed predefined search strings, a structured search procedure, predefined eligibility criteria, and manual verification by two authors of all extracted information, including all LLM assisted extractions, with disagreements resolved through consensus meetings. Finally, this study aimed to synthesize evidence from academic and practitioner-oriented sources to provide a comprehensive characterization of QST. Consistent with the objective of a multivocal literature review, we synthesized both evidence sources together rather than analyzing them separately. Consequently, the findings should be interpreted as an integrated view of the current QST ecosystem rather than as a comparative analysis between research and practice. In addition, as a preliminary multivocal literature review, this study does not claim to provide exhaustive coverage of all QST discussions, tools, frameworks, or practitioner communities.
\section{Conclusion and Future Work}

This paper presented preliminary findings from a multivocal literature review investigating how QST is characterized across both academic and practitioner-oriented sources. Our findings identified a rapidly growing and still fragmented ecosystem of testing techniques, challenges, tools, and infrastructures associated with QSE. Yet, these findings remain preliminary, and future stages of this research will incorporate additional evidence collected about testing recommendations, testing practices, and quality attributes addressed during quantum testing activities. Our future work will also separate academic literature from practitioner-oriented material to enable a more detailed comparative analysis of how research and practice converge, diverge, or complement one another in discussing QST challenges, techniques, and engineering needs.

\begin{databox}
\textbf{Data availability statement:}
To ensure verifiability and replicability, we provide the replication package of this study available at~\url{https://figshare.com/s/ce54db63291f2dea9ecd}.
\end{databox}

\bibliographystyle{IEEEtran}
\bibliography{biblio}

\end{document}